\documentclass[12pt]{article}
\usepackage{scicite}
\usepackage{times}

\usepackage[T1]{fontenc}
\usepackage{float}
\usepackage{graphicx}
\usepackage{pdflscape}
\usepackage{hyperref}
\usepackage{pgf,pgffor}
\usepackage{textcomp}
\usepackage{xcolor}
\usepackage{ulem}
\usepackage{lineno}
\usepackage{caption}
\usepackage[caption = false]{subfig}
\usepackage{pifont}
\usepackage{longtable}
\usepackage{amssymb}
\usepackage{cleveref}
\definecolor{orcidlogocol}{HTML}{A6CE39}
\usepackage[symbol]{footmisc}

\usepackage{deluxetable}

\newcommand{\apj}{Astrophys. J.}
\newcommand{\apjl}{Astrophys. J.}
\newcommand{\apjs}{Astrophys. J.}
\newcommand{\aap}{Astron. Astrophys.}
\newcommand{\mnras}{Mon. Not. R. Astron. Soc.}
\newcommand{\nat}{Nature}

\newcommand{\aapr}{Astronomy and Astrophysics Reviews}

\definecolor{orcidlogocol}{HTML}{A6CE39}

\makeatletter

\let\saved@includegraphics\includegraphics
\AtBeginDocument{\let\includegraphics\saved@includegraphics}
\renewenvironment*{figure}{\@float{figure}}{\end@float}
\makeatother

\makeatletter
\def\@fnsymbol#1{\ensuremath{\ifcase#1\or \dagger\or \ddagger\or
 \mathsection\or \mathparagraph\or \|\or **\or \dagger\dagger
 \or \ddagger\ddagger \else\@ctrerr\fi}}
\makeatother

\newcommand{\FIG}[1] {Fig.~\ref{#1}}
\newcommand{\TAB}[1] {Tab.~\ref{#1}}
\newcommand{\EXTTAB}[1] {Tab.~S\ref{#1}}
\newcommand{\EXTFIG}[1] {Fig.~\ref{#1}}
\newcommand{\AFF}[1]{$^{\foreach\d[count=\ni]in{#1}{\ifnum\ni=1\ref{\d}\else,\ref{\d}\fi}}$}
\definecolor{dkblue}{RGB}{54, 86, 169}
\newcommand{\REV}[1] {{\textcolor{black}{{#1}}}}

\newenvironment{sciabstract}{%
\begin{quote} \bf}
{\end{quote}}

\title{A radio pulsar phase from SGR J1935+2154 provides clues to the magnetar FRB mechanism }

\author{
Weiwei Zhu$^{*1~2}$,
Heng Xu$^{*3~1~5}$, 
Dejiang Zhou$^{*1~4}$, 
Lin Lin$^{2~6}$,  
Bojun Wang$^{3~1}$, \\
Pei Wang$^{1~2}$,  
Chunfeng Zhang$^{3~1}$,  
Jiarui Niu$^{1~4}$, 
Yutong Chen$^{1~4}$, 
Chengkui Li$^{7} $,\\
Lingqi Meng$^{1~4}$, 
Kejia Lee$^{3~1~5}$,
Bing Zhang$^{8~9}$,
Yi Feng$^{10}$, 
Mingyu Ge$^{7}$,\\
Ersin.~G\"{o}\u{g}\"{u}\c{s} $^{11}$, 
Xing Guan$^1$,
Jinlin Han$^1$, 
Jinchen Jiang$^{3~1}$, 
Peng Jiang$^{1~12}$,\\
Chryssa Kouveliotou$^{13}$, 
Di Li$^{1~10}$, 
Chenchen Miao$^{1~4}$, 
Xueli Miao$^1$,
Yunpeng Men$^{14} $, \\
Chenghui Niu$^{1} $, 
Weiyang Wang$^{4~3} $, 
Zhengli Wang$^{15}$ , 
Jiangwei Xu$^{3~1} $, 
Renxin Xu$^2$,\\
Mengyao Xue$^{1} $, 
Yuanpei Yang$^{16} $,
Wenfei Yu$^{17}$, 
Mao Yuan$^{1~4} $, 
Youling Yue$^{1}$, \\
Shuangnan Zhang$^{7} $,
Yongkun Zhang$^{1~4} $,\\ 
 \footnotesize{$^1$National Astronomical Observatories, Chinese Academy of 
 Sciences; Beijing 100101, China.}\\ %\footnotesize{zhuww@nao.cas.cn,\href{https://orcid.org/0000-0001-5105-4058}{\textcolor{orcidlogocol}{} \hspace{2mm} orcid.org/0000-0001-5105-4058}} \\
  \footnotesize{$^2$Institute for Frontiers in Astronomy and Astrophysics, Beijing Normal University, Beijing 102206, China.} \\
 \footnotesize{$^3$Department of Astronomy, Peking University; Beijing 100871,
 China.} \\
 \footnotesize{$^4$University of Chinese Academy of Sciences, Chinese Academy of Sciences; Beijing 100049, China.} \\
 \footnotesize{$^5$Kavli institute for astronomy and astrophysics, Peking University; Beijing. 100871, China.} \\
 \footnotesize{$^6$Department of Astronomy, Beijing Normal University; Beijing 100875 , China.} \\
 \footnotesize{$^7$Key Laboratory of Particle Astrophysics, Institute of High} \\
 \footnotesize{Energy Physics, Chinese Academy of Sciences; Beijing 100049, China.} \\
 \footnotesize{$^8$Nevada Center for Astrophysics, University of Nevada; Las Vegas, NV 89154, USA. }\\ %\footnotesize{bing.zhang@unlv.edu,\href{https://orcid.org/0000-0002-9725-2524}{\textcolor{orcidlogocol}{} \hspace{2mm} orcid.org/0000-0002-9725-2524} }  \\
 \footnotesize{$^9$Department of Physics and Astronomy, University of Nevada, Las Vegas; NV 89154, USA. } \\
 \footnotesize{$^{10}$Zhejiang Lab; Hangzhou, Zhejiang 311121, 
 China.} \\
 \footnotesize{$^{11}$Sabanc{\i} University, Faculty of Engineering and Natural Sciences, 34956; \.Istanbul, Turkey.} \\
  \footnotesize{$^{12}$Guizhou Radio Astronomical Observatory, Guizhou University, Guiyang 550025, China;}\\
 \footnotesize{$^{13}$Department of Physics, The George Washington University;}\\
 \footnotesize{725 21st St. NW, Washington, DC 20052, USA. } \\
 \footnotesize{$^{14}$Max-Planck Institut f{\"u}r Radioastronomie; Auf dem
 H{\"u}gel 69, D-53121 Bonn, Germany. } \\
 \footnotesize{$^{15}$GuangXi Key Laboratory for Relativistic Astrophysics,}\\
 \footnotesize{School of Physical Science and Technology; GuangXi University, Naning 530004, China. } \\
 \footnotesize{$^{16}$South-Western Institute for Astronomy Research, Yunnan} \\
 \footnotesize{University; Kunming 650500, Yunnan, China. } \\ 
 \footnotesize{$^{17}$Shanghai Astronomical observatory, Chinese Academy of Science; Shanghai 200030, China. } \\
 \footnotesize{$^{*}$These authors contributed equally}\\
 \footnotesize{Corresponding author. Email:zhuww@nao.cas.cn; kjlee@pku.edu.cn; bing.zhang@unlv.edu }
}

\date{}

\begin{document}
%\linenumbers
\baselineskip24pt
\maketitle

\begin{sciabstract}
The megajansky radio burst, FRB~20200428, and other bright radio bursts detected from the Galactic source SGR~J1935$+$2154 suggest that magnetars can make fast radio bursts (FRBs), but the emission site and mechanism of FRB-like bursts are still unidentified.
Here we report the emergence of a radio pulsar phase of the magnetar five months after FRB~20200428. 795 pulses were detected in 16.5 hours over 13 days by the Five-hundred-meter Aperture Spherical Radio telescope, with luminosities about eight decades fainter than FRB~20200428.
The pulses were emitted in a narrow phase window anti-aligned with the X-ray pulsation profile observed by the X-ray telescopes.
The bursts, conversely, appear in random phases.
This dichotomy suggests that radio pulses originate from a fixed region within the magnetosphere, but bursts occur in random locations and are possibly associated with explosive events in a dynamically evolving magnetosphere.
This picture reconciles the lack of periodicity in cosmological repeating FRBs within the magnetar engine model.

\end{sciabstract}

%\section*{Introduction}
% 

\section*{INTRODUCTION}
On 2020 April 28, a megajansky radio burst composed of two pulses was detected from a Galactic magnetar, SGR~J1935$+$2154\cite{CHIMEJ1935,STARE2J1935,LiCK2021,Ridnaia21,Tavani21,Mereghetti20}. This event, now called FRB~20200428, sheds light on the mysterious, cosmological fast radio bursts (FRBs) prevailing in the radio sky\cite{Petroff19AAR,Cordes19ARAA,ZhangB2020,Petroff22} and suggests that at least some FRBs originate from magnetars.
However, most magnetars are radio-quiet and occasionally emit radio emission that is highly variable \cite{camilo06,camilo07}.
The physical conditions and the mechanism for magnetars to emit bright coherent radio emission are still being debated \cite{Katz2018,Lyutikov2021,Lyubarsky2021,WWY2022,Thompson2023}.
Here we report an extended, deep follow-up radio observation of SGR~J1935$+$2154, which revealed the appearance and disappearance of a radio magnetar phase five months after the FRB burst. 
The radio pulses clearly show a period of 3.2478~s, with all the pulses arriving within a 7\% window in terms of the normalized pulse phase (\FIG{fig:waterfall}).
The phase of the radio pulses is anti-aligned with the X-ray pulsation phase, contrary to the FRB~20200428 event\cite{Younes20}.
The single pulses show complex sub-structures that 
resemble those observed from repeating FRBs.
A slow down of the pulsar rotation is clearly detectable and can be modeled with a steadily spinning-down magnetar with slow pulse profile evolution.

%Observations
\section*{RESULTS}
\begin{figure}[H]
 %\phascentering
 \centering
 \includegraphics[width=4.75in]{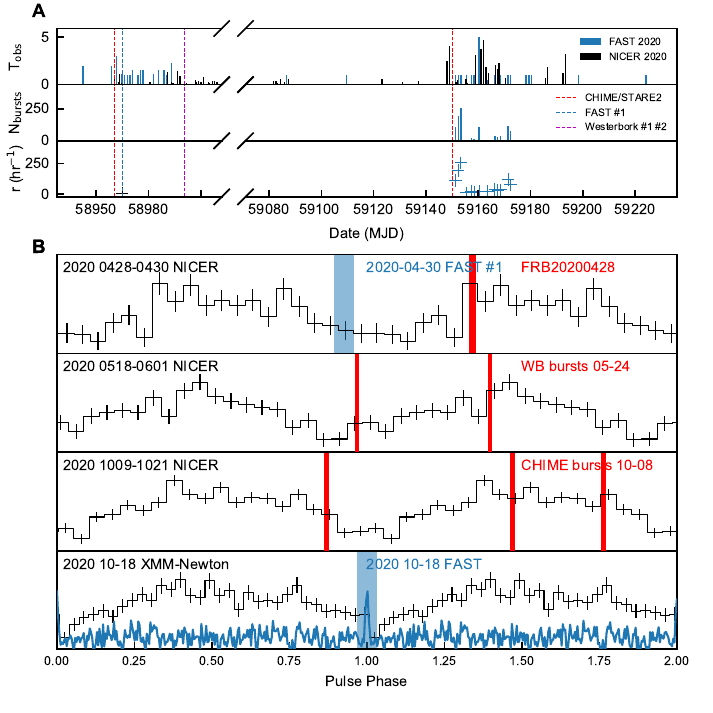}
 \caption{\textbf{The radio and X-ray campaign and the bursts' rotational phases.} {\bf A} The timeline of our FAST  observational campaign of SGR~J1935$+$2154 in cyan and NICER observations in black, with important events labelled with coloured vertical dash lines. $T_{\rm obs}$ is the FAST or NICER on-source observation time, $N_{\rm bursts}$ is the number of bursts detected by FAST, and $r$ is the FAST event rate in the unit of $1/{\rm hr}$.
{\bf B} Black curves represent the persistent pulse profile detected by NICER or XMM-Newton. 
We aligned the X-ray profiles across different panels by fitting them with sinusoidal curves except for the profile from April 28-30, 2020.
The NICER profile of April 28-30, 2020, differs in shape from the rest X-ray profiles, so we manually moved it by $+$0.2 to better align its peak with the other X-ray profiles.  
The vertical lines label the phases of the FRB~20200428 bursts (red), the FAST \#1 (cyan), the Westerbork bursts (magenta), the CHIME Oct 8 bursts (red) and the FAST pulsar radiation taken on Oct 18th (cyan curve).
We also present the FAST integrated profile from Oct 18th in cyan in the bottom plot of panel {\bf B}.\label{fig:spds}
}
\end{figure}

We monitored SGR~J1935$+$2154 using the Five-hundred-meter Aperture Spherical Telescope (FAST\cite{Jiang19SCPMA}) since 2020 April 15 even before the occurrence of FRB 20200428 and started a dense daily observation campaign from April 28 to May 19 with a total duration of 26 hours (\FIG{fig:spds} \textbf{A}, \EXTTAB{tab:Obs}). 
%Despite the discovery of a megajansky level FRB-like radio burst (i.e. FRB~20200428)\cite{CHIMEJ1935, STARE2J1935} by the Canadian Hydrogen Intensity Mapping Experiment (CHIME) and the Survey for Transient Astronomical Radio Emission 2 (STARE2), 
Even though FRB~20200428 was temporally associated with a hard X-ray burst \cite{LiCK2021,Ridnaia21,Tavani21,Mereghetti20},
no radio burst was detected by FAST when the source was emitting many other hard X-ray bursts \cite{LinZhang2020}. 
Two days after the FRB event, we caught one highly polarised isolated radio event on 2020 April 30 (UTC 2020-04-30T21:43:00.499) \cite{ZhangCF20} (denoted as ``FAST \#1'' hereafter) with an estimated duration of 0.93(4)~ms, a flux of 30.7(4)~mJy, and a fluence of 51(2) mJy~ms.
%Assuming a distance of 6.6~kpc\cite{zhou20}, the radio isotropic peak luminosity is 7.2(1)$\times10^{29}$~erg/s and the isotropic radio energy is 1.19(4)$\times10^{27}$~erg. 
This pulse is fainter than FRB~20200428 by a factor of 4$\times10^6$ in isotropic peak luminosity and by 7 orders of magnitude in isotropic energy. 
Due to the lack of radio activity, we stopped monitoring the source after May 19, but revisited twice in August, again without any radio emission detection. 
Continued monitoring with the Westerbork telescope in the Netherlands later detected two moderately bright radio bursts on 2020 May 24 with specific fluences 112(22)~Jy~ms and 24(5)~Jy~ms, respectively \cite{kirsten2021WB}. These events are rare, as most other monitoring campaigns conducted with the Arecibo, Effelsberg, Low Frequency Array (LOFAR), MeerKAT, Mark 2 (MK2), Parkes telescopes \cite{Bailes21,Tang21} have led to non-detections.

\begin{figure}[H]
 \centering
 \includegraphics[width=4.0in]{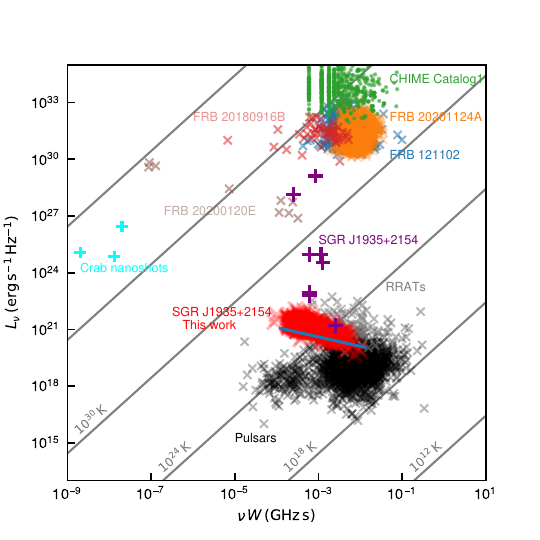}
 \caption{\textbf{Comparison of various SGR~J1935$+$2154 radio emissions with emission from other sources.} The y-axis is the radio luminosity, and the x-axis is the product of observing frequency and the width of radio emission (bursts or pulses). The thick red crosses denote the single pulses of SGR~J1935$+$2154 reported in this work. The blue line indicate the position of the detection threshold for the FAST observations. The radio bursts from SGR~J1935$+$2154 are shown in purple pluses, from top to bottom are the FRB-like burst detected by CHIME\cite{CHIMEJ1935}, STARE2\cite{STARE2J1935}, intermediate bursts detected by Westerbork\cite{kirsten2021WB} and CHIME\cite{CHIME20ATel}, and pulse \#1 detected by FAST\cite{ZhangCF20}. Emission from other active FRB sources, FRB~20201124A\cite{Xu2022}, FRB~121102\cite{Li2021}, FRB~20180916B\cite{Nimmo20180916B,CHIME2020periodicity}, FRB~20200120E\cite{Nimmo20200120E}, and FRBs from CHIME catalogue1\cite{CHIMECatalog1} are plotted in different colors. The emission from pulsars and Rotating Radio Transients are shown in grey crosses and pluses, respectively, with the cyan pluses representing Crab nanoshots\cite{Hankins2003, Hankins2007, Jessner2010}. }
 \label{fig:phasespace}
\end{figure}

On 2020 Oct 8, the Canadian Hydrogen Intensity Mapping Experiment (CHIME) telescope detected three moderate-luminosity bursts with specific fluences of 900(160), 9(2), 6(1)~Jy ms, separated by 1.949~s and 0.954~s between adjacent bursts, respectively\cite{CHIME20ATel}.
We promptly re-started our monitoring campaign on 2020 Oct 9 (UTC), using FAST with the center beam of its 19-beam receiver in the a band centered at 1250~MHz\cite{Jiang20RAA}.
%single pulses and pulse evolution
We observed the source for 1-5 hours per day for 13 days in 2020 October (spanning from 9th to 30th), and detected a total of 795 pulses that showed clear periodicity.
%Energy
%The pulsar phase of SGR J1935$+$2154 is drastically different from the FRB phase in terms of radio pulse energy. 
The magnetar's distance was estimated to be about 9~kpc \cite{sun11} or 12.5~kpc \cite{kothes18} by its association with the supernova remnant G57.2+0.8, however, Ref.
%Zhou et al. (2020)
\cite{zhou20} revisited the supernova remanent (SNR) distance by using CO line observations and estimated its distance to be $d \sim 6.6~{\rm kpc}$.
Later, Ref. \cite{Bailes21} examined the SNR distance by combining multi-frequency observations and argued that it is even closer $d=1.5-6.5~{\rm kpc}$.
In this paper, we assume a distance of $d=6.6~{\rm kpc}$ since this value bridges the above estimations.
The luminosities of single pulses are in the range of 1.8$\times10^{28}$~erg~s$^{-1}$ to 4.5$\times10^{30}$~erg~s$^{-1}$.
If the magnetar is indeed much closer than assumed here, then the estimated burst luminosities and energies would have also been lower.
As shown in \FIG{fig:phasespace}, the single pulse luminosities are about 8-9 orders of magnitude lower than that of FRB~20200428\cite{CHIMEJ1935,STARE2J1935} or 3 orders of magnitude lower than those of the intermediate bursts\cite{kirsten2021WB,CHIME20ATel}.
Assuming a distance 6.6~kpc, we find that the single pulse luminosities are consistent with those of Rotating Radio Transients (RRATs). With a smaller distance \cite{Bailes21}, they fall into the range of radio pulsars.
%By converting the total single pulse energies observed in our observing time (when the pulsar is active) into an average radio luminosity, 
We find that the average single pulse \REV{luminosity} $6.2(2)\times10^{29}$erg~s$^{-1}$ is about 10$^{-5}$ of the pulsar's spin down luminosity ($\dot{E}=4.28\times10^{34}$erg~s$^{-1}$) inferred from the timing of the single pulses. \REV{Since} the pulsed radio emission is short and intermittent, the total observed pulse energy is about 10$^{-10}$ that of the spindown energy when averaged over the observation time. 
This is consistent with the suggestion \cite{Rea2012} that magnetar pulsed radio emission is powered by the spindown of the magnetar.
\REV{Conversely, at a distance of 6.6~kpc, the FRB~20200428 event consists of a peak luminosity of $3(3)\times 10^{36}$~erg~s$^{-1}$ \cite{CHIMEJ1935}, substantially exceeding the pulsar's spindown power, suggesting a different energy source and mechanism are required for the FRB-like bursts.
}

\begin{figure}[H]
 \centering
 \includegraphics[width=4.5in]{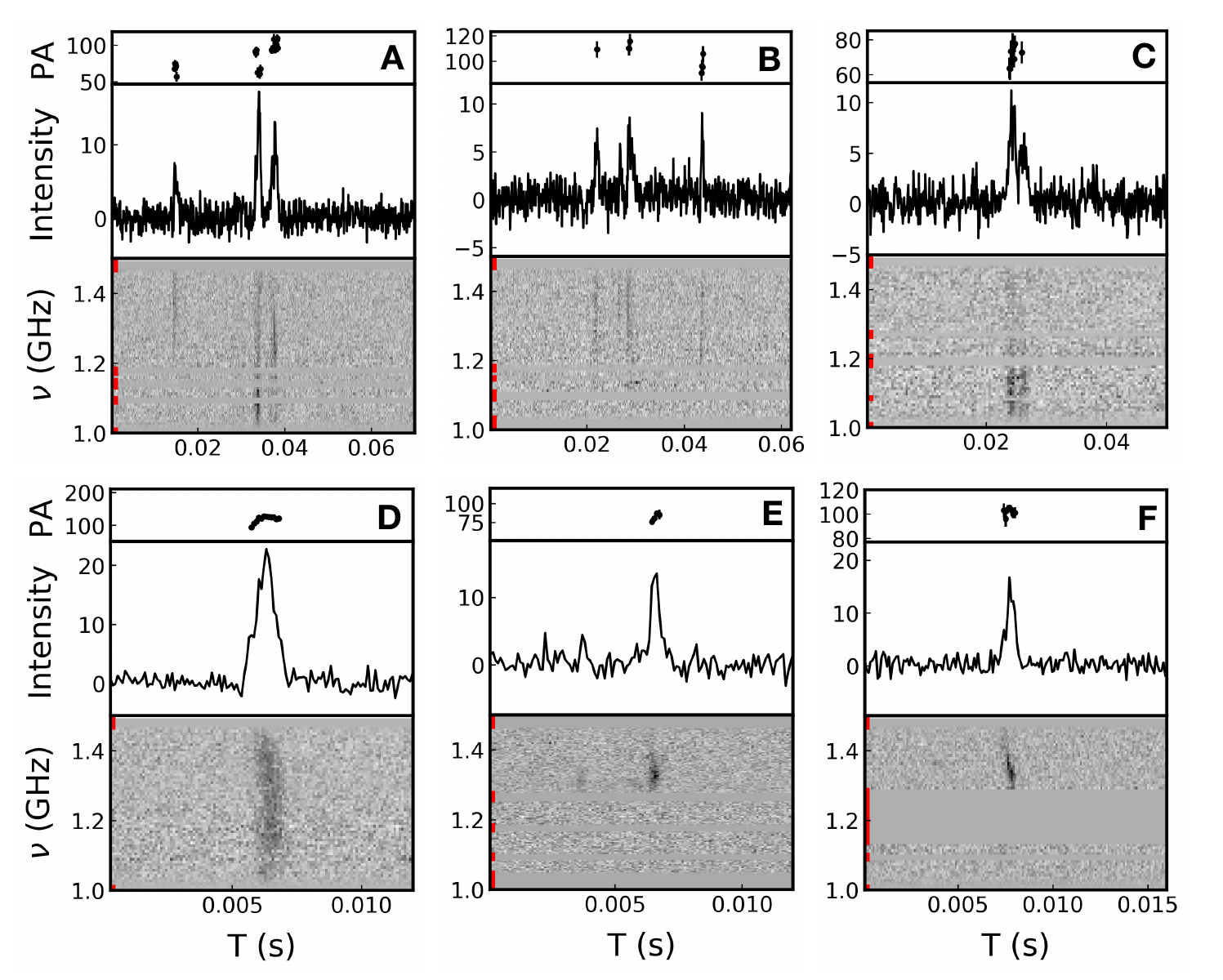}
 \caption{\textbf{A gallery of emission properties of SGR~J1935$+$2154's pulses.} The x-axis $T$ is time in units of seconds. We present in each panel the polarisation angle (PA) at 1250~MHz in units of degree, the pulse intensity profile with its off-pulse variance normalized to unity, and the gray-scale image display the power as a function of both time $T$ and frequency $\nu$ (also called the ``waterfall'' plot) after dedispersion of radio signals with DM=332.703 cm$^{-3}$~pc.
 Panels \textbf{A B C} show pulses with multiple components. Panels \textbf{D E F} show pulses with narrow bandwidth and frequency drift features.
 }
 \label{fig:multipeaksexample}
\end{figure}

The FAST-detected pulses are narrow with a mean Gaussian width of $\sim$0.9~ms. % (\EXTFIG{fig:width}).
Upon inspection of the FAST single pulses, we found pulses were detected from 399 individual pulsar rotations. We found multi-peak fine pulse structures or multiple pulses from 258 of these rotations.
These pulses or fine structures are separated by few-to-ten milliseconds, and some show narrow-band features and frequency drifts (\FIG{fig:multipeaksexample}).
These are considered some of the predominant features of repeating FRBs\cite{Hessels19, Li2021, Xu2022}, although some similar patterns have also been found from a small group of pulsars \cite{Bilous21} in low frequencies. 
%DEL{The resemblance between the single pulses of this FRB-generating magnetar and those of repeating FRBs indicates that the magnetar pulses might be emitted with a similar mechanism as the repeating FRBs.}
\REV{The resemblance between the single pulses of this FRB-generating magnetar and those of repeating FRBs indicates that the plasma responsible for the two types of coherent bursts might share some physical traits which are imprinted in the dynamic spectrum even though the physical mechanism behind their creating might be completely different.}

\begin{figure}[H]
 %\phascentering
 \centering
 \includegraphics[width=4.75in]{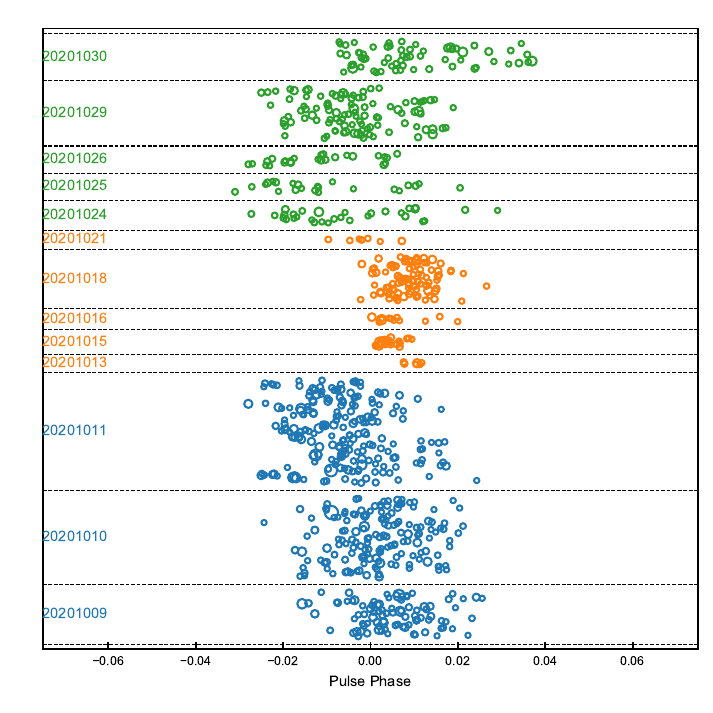}
 \caption{\textbf{ A plot of spin phases of the radio pulses detected by FAST in October 2020}. 
 The size of the marker represents the S/N of the pulse.
 The phases of the pulses are computed based on our best fit timing solution (\TAB{tab:parfile}).
 We separate the emission into three phases (marked by blue, orange, and green, respectively), of which the pulse phase distribution is different.\label{fig:waterfall}
}
\end{figure}

A periodicity is detected using the single pulse timing technique (Supplementary material). 
After folding the single pulse arrival times and the dedispersed data with the coherent timing model in Tab. \ref{tab:parfile}, both the single pulses arrival times and the folded profile peaks fall in a phase window of 7\%.
The small phase spread corresponds to an observed opening angle of the emission region $W=25.2^{\rm o}$. 
The pulse-arriving phases are slowly varying with time. During Oct 13-21, the pulse phase spread became as small as $\sim 2$\% (the yellow cycles in \FIG{fig:waterfall}) and relaxed to $\sim 5$\% after Oct 24 (the green cycles in \FIG{fig:waterfall}).  
%The pulse phase evolution implies a varying emission region resulting from a slowly evolving magnetosphere. 
The pulse phase evolution could be the result of either a low level spin noise or the variations of the emission height or field lines at different emission episodes, probably related to the varying pair production conditions near the surface region. The latter scenario could explain both the phase evolution and the associated phase spread changes.
Such a pulse duty cycle implies an emission region of at least a hundred times of the stellar radius within the magnetar magnetosphere unless the magnetic axis is nearly aligned with the spin axis.
\REV{Note that the pulsar's X-ray emission is consistent with being from the surface hot spots heated by the return currents from the magnetic pole \cite{Borghese22}.
In the scenario of an aligned rotator, the hot spots would remain almost stationary as the pulsar spins, producing little pulsation, which is further reduced by the gravitational lensing effect of the neutron star.
However, the observed X-ray pulse profile shows a consistent shape and varies by more than a factor of two in counts per bin in all four sessions taken months apart. }
Therefore, a nearly-aligned magnetic axis is incompatible with the observed X-ray pulsation, leading us to believe that the emission region of the pulses must be much higher in the magnetosphere.

\begin{figure}[H]
 \centering
 \includegraphics[width=6.0in]{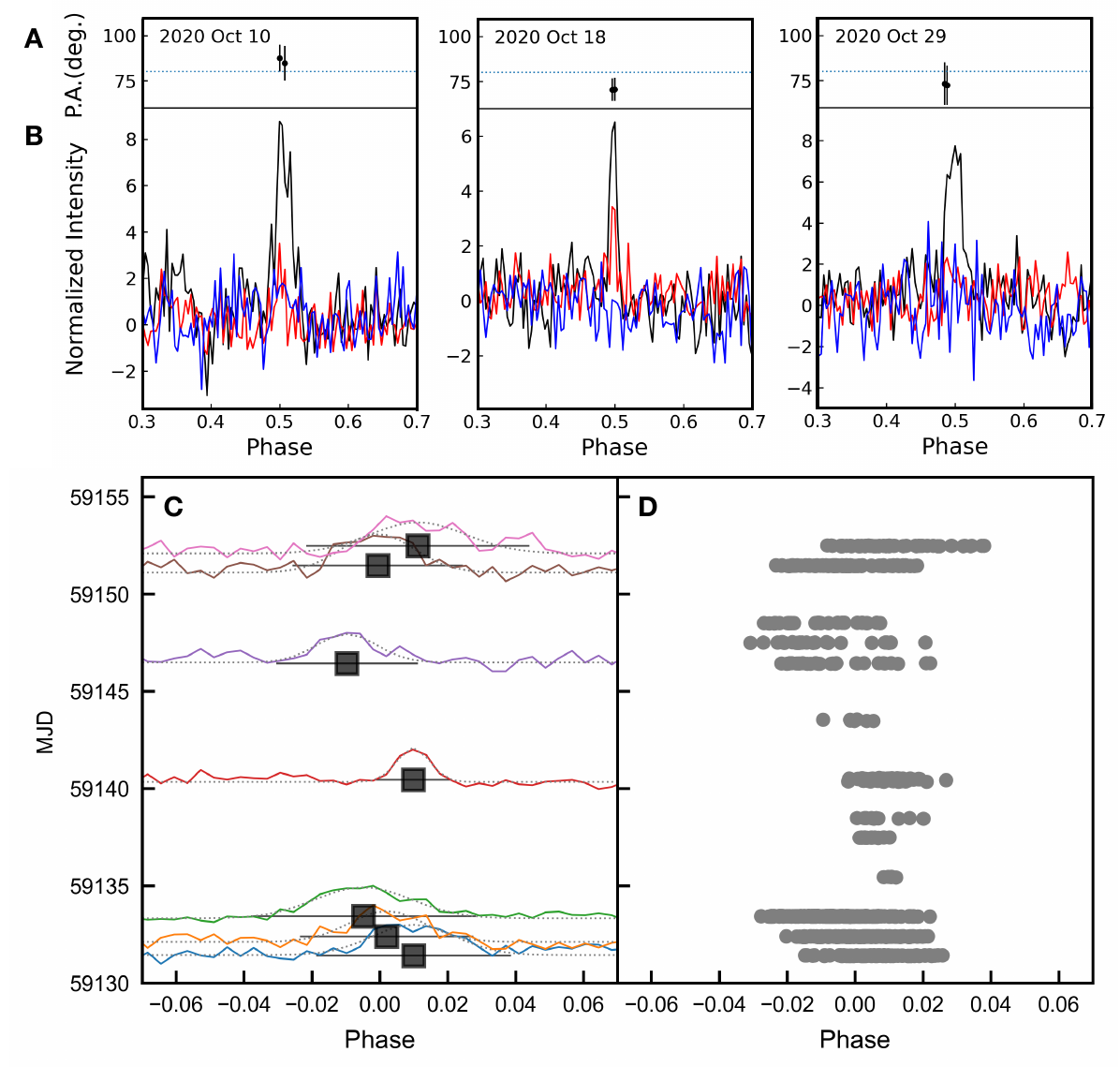}
 \caption{
 \textbf{Folded pulse profiles on 2020 Oct 10, 18 and 29 and the evolution of the folded profiles and single pulses.} \textbf{A} Polarisation position angles measured at 1250~MHz. \textbf{B} Polarisation integrated pulse profile, with total intensity, linear polarisation, and circular polarisation in black, red, and blue, respectively.
 \textbf{C} folded pulse profiles (coloured curves). 
 The horizontal box error bars show the center and the 90\%-flux width of the best fit Gaussian. 
 \textbf{D} Phases of the single pulses.}

 \label{fig:profile}
 \label{fig:profevo}
\end{figure}

\begin{figure}[H]
 \centering
 \includegraphics[width=4.0in]{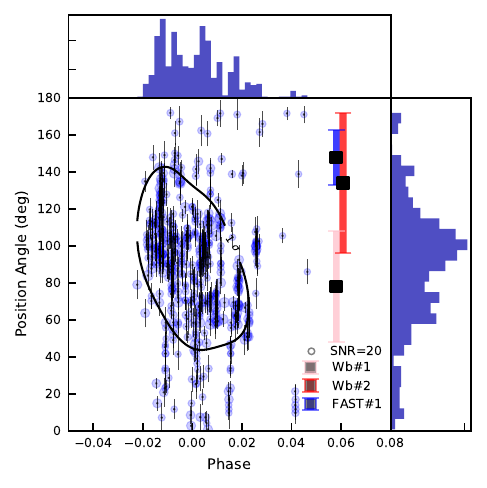}
 \caption{
\textbf{Polarisation position angle-pulse phase distribution of bright single pulses (S/N>20). }
We compare the polarisation position angles of the Westerbork bursts (red bars) and that of the FAST pulses (blue bars) after derotating them to infinite frequency. 
Note that the Westerbork bursts (Wb \#1, Wb \#2) and the FAST \#1 burst are placed in arbitrary phases for convenience of comparison. We cannot extend our timing solutions to compute the phases of those bursts accurately.
For the FAST pulses we assumed a fixed RM of 111~rad~m$^{-2}$.
The size of each FAST data point is proportional to the S/N of the pulse, and the reference marker of S/N=20 is plotted. 
The solid contour labeled with 1-$\sigma$ is the 68\% contour of the cumulative distribution of FAST PAs derived from a Gaussian kernel density estimation\cite{scott2015multivariate}.
\label{fig:PAdist}}
\end{figure}

Both the individual pulses and the integrated pulse profile (detectable in a few  epochs) are polarized.
From bright pulses (S/N>20), we find a weighted-average rotation measure (RM) of 111(4)~rad~m$^{-2}$, which is consistent with that derived from pulse \#1\cite{ZhangCF20}.
The RMs of bright pulses are consistent with this average value with a reduced $\chi^2$ of 1.68.
Therefore we use a fixed RM of 111~rad~m$^{-2}$ for our analysis.
The polarisation position angle (PA) for bright pulses and the folded profiles are presented in \FIG{fig:multipeaksexample}, \EXTFIG{fig:profile}~\textbf{A} and \EXTFIG{fig:PAdist}. 
The single pulse PAs (at infinite frequency) show a broad Gaussian %symmetrical 
distribution centered around 90$^{\circ}$ from west to north.
The integrated profile has a linear polarisation degree $\Pi$ between 0.12(4) and 0.65(5).
The PA of the folded pulse (centered at 1250~MHz) stays roughly constant around $\sim80^{\circ}$ across phase. There is no sign of substantial PA swing in the folded profile \EXTFIG{fig:profile}~\textbf{A}, suggesting that the line of sight may be grazing the emission beam, putting their emission region even higher than the estimation based on the pulse profile width. 

Our timing solution contains a derived period 
$P=3.24781628(3)$~s and period derivative $\dot P = 3.717(3)\times10^{-11}$~s~s$^{-1}$ at a reference epoch of modified Julian date (MJD) 59131.377113 (\TAB{tab:parfile}),
corresponding to a derived surface magnetic field at the equator of about $3.52\times10^{14}$~G and a characteristic age of about $1.38\times10^{3}$~yr.
The spin down rate differs substantially from the $1.43(1)\times10^{-11}$~s~s$^{-1}$ value measured for this magnetar between 2014 and 2015 by Israel et al. \cite{israel16} (at reference MJD 56866) using the Chandra and XMM-Newton observations, and is closer yet still different from the value $3.92(2)\times10^{-11}$~s~s$^{-1}$ measured from the Neutron Star Interior Composition Explorer (NICER) observations\cite{Younes20} taken between 2020 April and May (at reference MJD 58997.571), indicating a notable pulsar spin evolution over months to years between 2014 and 2020. 
Such a noisy timing behavior is not uncommon among active magnetars.

The single pulse timing residuals show clear temporal evolution. 
The folded pulse profiles were also detectable ($\ge 5\sigma$) for some of the days (\EXTFIG{fig:profevo}~\textbf{B} and \textbf{C}).
The 90\%-flux pulse duty cycle was 0.051(6) to 0.066(4) in 2020 Oct 9 - 11 and dropped to 0.022(3) on Oct 18 accompanied by a center phase offset of 0.015(1). The duty cycle rose back to 0.04(2) to 0.066(9) between Oct 24 to Oct 30 with a center phase offset of -0.0196(2) from Oct 18. 
We triggered simultaneous XMM-Newton (9~hr) and FAST (5~hr) observations on Oct 18th, with the latter 5~hr observations overlapping with each other.
We compare the Oct-8 integrated radio and X-ray pulse profiles in \FIG{fig:spds}~\textbf{B}, which shows a $186(9)^\circ$ phase offset.
During the radio phase shifts from Oct. 13 to 18 and on Oct. 30, no obvious X-ray phase shift was observed\cite{Younes22}. These shifts are therefore unlikely caused by crustal movement on the neutron star surface, but is more likely related to magnetospheric reconfiguration during the period, suggesting a dynamically evolving magnetosphere of a radio-active magnetar. 
Similar profile evolution was also observed from other magnetars (e.g. Swift J1818.0-1607, XTE J1810-197) during their radio active phases \cite{rajwade22, maan22, caleb22}.
Finally, we conducted {\it Insight}-HXMT\cite{HXMT} observations of the source on Oct 10, 11, 13, 15, and 16, but did not detect any X-ray burst during the period.

Meanwhile, SGR~J1935$+$2154's surface thermal emission were detectable by X-ray telescopes such as NICER and XMM-Newton (\FIG{fig:spds}~\textbf{B}). The first NICER observation started hours after the source was reported to be in outburst by other wide-field space instruments and about 13.9 hours preceding the FRB~20200428 event. The observation caught many short X-ray bursts and the decay of the source from its extreme active state but unfortunately missed the exact moment when the FRB went off\cite{Younes20}. 
The X-ray pulsations provide a way to track the rotational phase of the star.
Ref. \cite{Younes20} compared the phase of FRB~20200428 and the Westerbork bursts with NICER persistent profiles of similar epochs and found that the phase of FRB~20200428 aligns with the peak of the X-ray pulse, while the two Westerbork bursts from 26 days later do not follow the same pattern. 
Later, CHIME detected three intermediate luminosity bursts again in Oct 8, 2020.
We find that the phases of the three CHIME bursts are, again, randomly distributed with respect to the X-ray profile from nearby epochs. 
These measurement show that the bursts are not confined to a specific rotational phase of the magnetar, but are random and sometimes separated by nearly a hundred degrees.
Furthermore, the aforementioned bursts all happened within one rotation period of the pulsar. This indicates that the driving mechanism behind the bursts does not persist for longer than seconds.

\section*{DISCUSSION}
%FAST started monitoring SGR~J1935$+$2154 on Oct 9, 2020 and detected weak periodic pulses from the magnetar.
By comparing with the X-ray profile detected by NICER and XMM-Newton, we find that the radio-to-X-ray phases of these FAST-detected pulses are aligned with the foot of the X-ray profile (\FIG{fig:spds}~\textbf{B}). 
More interestingly, when we compare the FAST \#1 pulse from April 30, two days after the FRB event, to the NICER profile from April 28-30, the FAST radio pulse appear also anti-aligned with X-ray profile.
This clear distinction between the bursts and the pulses, combined with their 3-8 orders of magnitude luminosity difference, suggests that the FRB-like bursts and their lower-luminosity brethren likely originate from drastically different emission sites from the pulsar mode pulses, and probably caused by different mechanisms.
We propose that the bursts are emitted when strong magnetosphere disturbances happen, possibly associated with some explosive processes. Multiple adjacent bursts may be related to different disturbances, or when the same disturbance travels in different field lines and make radio emission in different locations in the magnetosphere.
%or when a disturbance travels along the open or closed field lines. 
%Suppose the initial disturbance originates from short explosive events, such as glitches in the neutron star crust. 
For the latter scenario, multiple bursts could happen within one rotational period of the magnetar, which is what is observed. 
%, i.e. the time for any disturbance to sweep through and leave the magnetosphere.
Radio pulses, conversely, originate from a fixed region in the magnetosphere (probably the open field line region) and are powered by the standard radio pulsar mechanism. 
This picture could also explain why we have not yet observed any spin period from active repeating FRBs \cite{Li2021, Katz2022, Xu2022, NiuJR2022, Jahns2022} because the FRB bursts were emitted from a temporarily and dynamically distorted magnetosphere. 
Continued monitoring of SGR~J1935+2154 and other Galactic magnetars in both radio and X-ray bands will further test this ansatz of the dichotomy of radio emissions from magnetars.  

%\section*{Materials and Methods}
\section*{MATERIALS AND METHODS}
\subsubsection*{FAST observations}
We observed SGR J1935$+$2154 for 40 sessions using the center beam of the FAST 19-beam receiver\cite{nan2011five, Jiang19SCPMA} which operates in the 1.0-1.5~GHz range between 2020 April 15 and 2021 February 4 (\EXTTAB{tab:Obs}). 
The FAST data were recorded with the re-configurable open architecture computing hardware version 2 (ROACH2) System \cite{hickish2016decade} 
in the AA, BB, AB, BA four-polarisation channel PSRFITS format. The spectral resolution and time resolutions are 112.07~kHz and 98.304~$\mu$s, respectively.
A 1~minute winking noise of 10~K-equivalent power and 0.2 second period was injected at the beginning of the observation for polarisation calibration of the data.

\subsubsection*{Radio pulse search}

We combined the AA and BB channels to form the 8-bit intensity data with software package \textsc{psrfits\_utils} and automatically adjusted the level of the data in the process.
The level adjusting procedure rescales the level of the combined intensity according to its running standard deviation and also flattens the bandpass (the time-averaged intensity as a function of frequency), except for parts of the band that are heavily contaminated by interference.
We used the tool \textsc{rfifind} in the software package \textsc{PRESTO}\cite{Ransom11}  to search for the instantaneous and narrow-band radio frequency interference (RFI) and generate the RFI masks. 
The RFI masks were then combined with a list of bad channels, which were either known to be contaminated by the satellite signals or picked up visually according to their impact to the averaged bandpass. 
We used the tool \textsc{prepfold} in \textsc{PRESTO} to dedisperse the data while mitigating the RFIs using the above RFI information. 
We used 64 trial Dispersion measures (DMs) uniformly distributed between 328 and 337~pc~cm$^{-3}$ for the dedispersion. Such a fine DM grid ensures 3\% maximal S/N loss for the DM mismatching\cite{Men19MN}. 
The DM values of the detected pulses fall well within our searched range and the DMs distribute narrowly between 330.90 and 334.6~pc~cm$^{-3}$ with a weighted averaged value of 332.703(3)~pc~cm$^{-3}$, which is used in the later analysis.
After dedispersion, we used the \textsc{single\_pulse\_search.py} tool to search for single pulses with a minimum sigma cutoff of 5.5 and a list of box car width of 1, 2, 3, 4, 6, 9, 14, 20, 30, 45, 70, 100, 150, 220, 300 samples (maximum box car width of 29.5~ms). 
We inspected the search results by eye and selected pulses showing visible wide-band ($>50$~MHz) pulse.
We searched down to S/N$>5.5$, because of three reasons. First, the search is limited to a narrow range of DM, so the false positives drop notably. Second, since the pulse is narrow ($\sim 1$ ms), the effect of correlated noise becomes less important\cite{ZXM21}. Third, the pulses are faint, some are detectable at the level of S/N$=5.5$.
We compared our results with those produced from other pipelines using the software package \textsc{TransientX} or \textsc{BEAR}\cite{Men19MN}.
The differences were mainly caused by how the software isolates the sub-pulses and the detected pulse population is compatible.

\subsubsection*{Single pulse timing}

In the pulsar timing procedure, we dedispersed all pulses using the best estimated DM of 332.703~pc~cm$^{-3}$. 
The pulses, mostly milliseconds in width, have a Gaussian function shape.
We treated each subpulse individually with minimal separation between them of 1~ms.
We then fitted the dedispersed profile of each pulse with a Gaussian function. We converted the best fit pulse arrival times into the terrestrial time of arrivals (TOAs) at the FAST site.
We used the width of the pulse divided by its S/N as the TOA error.
We used the FAST-site onsite clock correction file to correct the site time to UTC. 
We used software package \textsc{tempo}\cite{nds+15} and \textsc{tempo2}\cite{hem06} to fit the pulsar timing model from the TOAs. 
In the procedure, we used the position parameters (Right Ascension and Declination) of SGR J1935$+$2154 derived from the X-ray imaging and assumed a position error of 0.4"\cite{israel16}. 
This is because our data is not long enough to cover the Earth's annual orbit and provide an independent position for the pulsar. 
To incorporate the uncertainties of the X-ray position in the timing analysis, 
we fitted for the spin frequency $\nu$ and its derivative $\dot{\nu}$ in a Monte Carlo Markov Chain (8000 mixing steps and 40000 iterations) in which R.A. and Decl. are sampled around the X-ray position with a standard deviation of $\simeq$0.4".
The resulting timing models were expressed in Barycentric Dynamical Time (TDB) scale by using the solar system ephemeris DE436. 
The initial attempt yielded a fit with extraordinarily high $\chi^2$, showing that the TOA residuals are far more scattered than the TOA errors.
This is caused by the fact that the single pulses arrive in a range of phases that span about 230~ms (or 0.07 in phase).
We used an EQUAD of 37.9~ms to rescale the TOA error, and that brings the reduced $\chi^2$ of timing fit to unity.
Both \textsc{tempo} and \textsc{tempo2} provided consistent best fit timing ephemeris (\TAB{tab:parfile}).
This timing solution is well consistent with the solution derived from NICER data of the same epochs \cite{Younes22}.
While our data constrains the spin parameters to a better precision than the NICER one, our data does not cover the date range between Oct 2 and Oct 9.

We folded the data with the timing solution derived from the single pulse timing. Integrated pulse profiles were detected (S/N$\ge$5) on several days as indicated in \EXTFIG{fig:profile} panels \textbf{B} and \textbf{C}. 
The integrated profiles show polarisation PAs that are flat around 80$^{\circ}$ during the ``on'' pulse phase at 1250~MHz (\EXTFIG{fig:profile}~\textbf{A}).

\subsubsection*{Pulse fine structures and morphology}

\begin{figure}[H]
 \centering
 \includegraphics[width=5.0in]{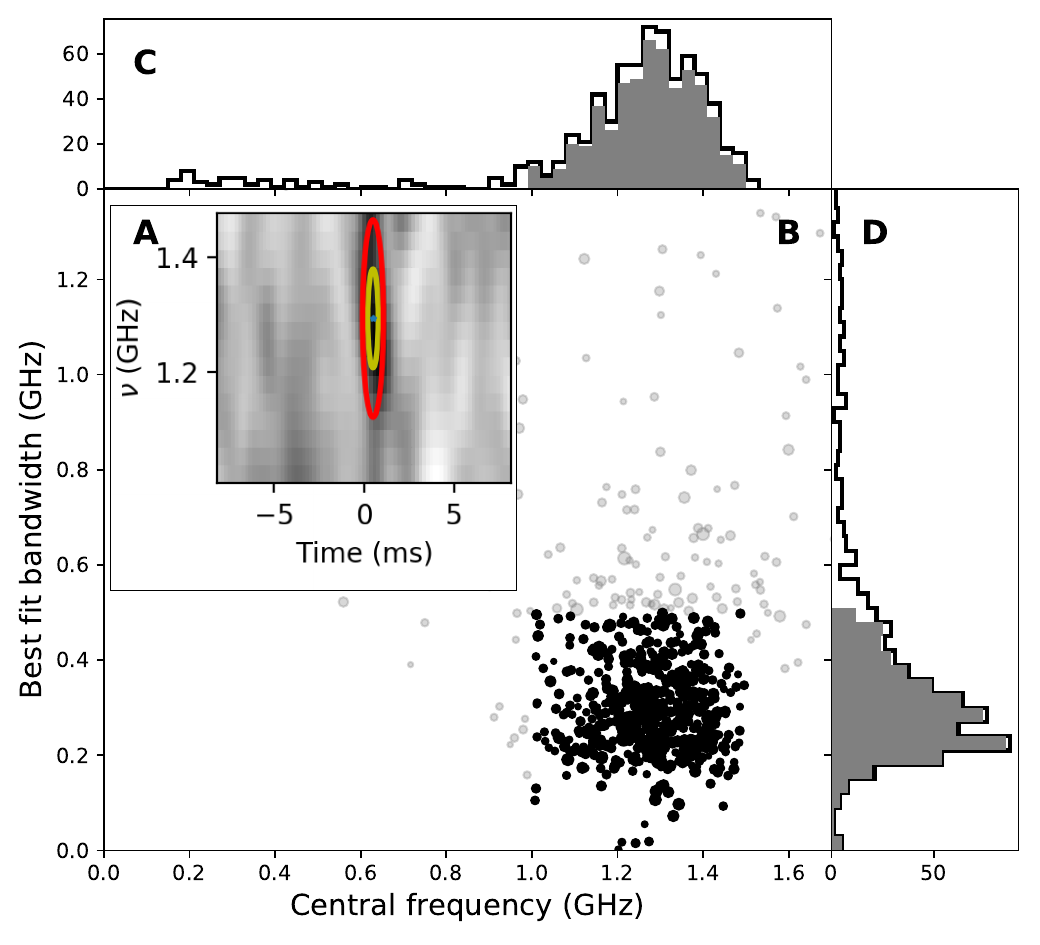}
 \caption{
 \textbf{Pulse bandwidth and center frequency distribution of SGR~J1935+1054.}
\textbf{A} A downsampled snapshot of a single pulse which is smoothed by using a Gaussian filter of a 2-pixel radius. The resulting smoothed image is fitted with a 2D vertical Gaussian profile (the yellow contour in the inlet marks the 1-$\sigma$ radius ellipse, and the red marks the 2-$\sigma$ radius). We take the center of the resulting vertical Gaussian profile as the central frequency of the pulse and two times the best fit Gaussian width (encompassing 95\% of the pulse energy) as the characteristic bandwidth for the single pulse. %\DEBUG{put the method text into Method section} 
\textbf{B} The distribution of the central frequency and bandwidth for the pulses. Some pulses' emission comes mainly from one side of the observing band, their central frequency could be outside the observable range, and their bandwidth could be larger than the observable bandwidth (450~MHz). In these cases, we plotted them using gray points.
In panel \textbf{C} we show two histograms, the step histogram shows the distribution of center frequencies for all pulses, and the filled histogram shows the pulses with center frequency inside of our observable range and bandwidth smaller than our observable bandwidth. 
Similarly, panel \textbf{D} shows the distribution of best fit bandwidths of all the pulses and pulses with reasonable fit parameters. 
 }
 \label{fig:bandwidth}
\end{figure}

To characterize the magnetar pulses' pulse width, emission band, and bandwidth, we analyzed all the pulses' positions and shapes and fit them with 2D Gaussian eclipses in the time and frequency space.
We took the center position in the frequency of the 2D Gaussian eclipse as the center frequency of the pulse and twice the Gaussian width (encompassing 95\% pulse energy) in frequency as the pulse bandwidth.
We found that the best fit central frequencies have a mean value of 1.28~GHz, a median of 1.29~GHz, and a standard deviation of 0.10~GHz, and the best fit bandwidths have a mean value of 0.29~GHz, a median of 0.28~GHz, and a standard deviation of 0.09~GHz (see \EXTFIG{fig:bandwidth}). 
These pulse characteristics are comparable to the narrow-band features seen from some repeating FRBs, such as FRB~121102\cite{Hessels19,Li2021}.
We also found that the Gaussian fit width of our detected pulses showed a median value of 0.5~ms, a mean of 0.9~ms, and a standard deviation of 0.9~ms;
while the equivalent width showed a median value of 0.5~ms, a mean value of 0.8~ms and a standard deviation of 0.8~ms. % (see \EXTFIG{fig:width}).

\begin{figure}[H]
 \centering
 \includegraphics[width=6.0in]{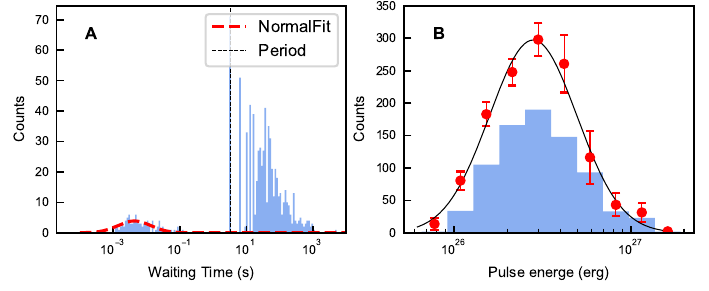}
\caption{\textbf{The waiting time and energy distribution of SGR~J1935$+$2154.} \textbf{A} The waiting time distribution. The vertical dotted line marks the period of the pulsar. We fit the millisecond-scale waiting times (red dotted curve) with a normal distribution in the logarithmic scale and find a peak at 5(1)\,ms with a $\sigma$ of 0.6(1) dex. 
\textbf{B} Energy distributions of the single pulses in the Oct 2020 episode 
filled histogram: the detected energy distribution of the SGR~J1935$+$2154's single pulses.
Red error bars: the estimated energy distribution after correcting for the sample completeness. 
Solid curve: the best-fit log-normal distribution curve.
}
 \label{fig:waitingtime}
 \label{fig:energy}
\end{figure}

\EXTFIG{fig:waitingtime}~\textbf{A} shows the distribution of waiting intervals, i.e., the time lapse between the best fit Gaussian center position of consecutive pulses. It shows two groups of intervals: 
One group of intervals is of millisecond time scales. They are the separations between multi-peak (subpulse) fine structures. 
These shorter waiting times show a peak at 5\,ms. 
The other group of intervals is of seconds time scales, and from the waiting time between consecutive pulses separated by at least one rotation. 
The tallest peak in this group coincides with one rotation period, 3.2478~s, and serials of harmonic periods follow it.
The longest waiting time reaches hundreds of rotations, displaying the intermittent nature of the emission.

%Fine structures. move from main text to here
\FIG{fig:multipeaksexample} shows a gallery of pulses from SGR~J1935$+$2154 detected by FAST in October 2020. 
We find a variety of fine pulse structures that resembles those observed from repeating FRBs\cite{Hessels19, CHIMEspectra, Li2021, Xu2022}.
Panels \textbf{A-C} show examples of multiple pulses arriving with separations of 5-10~ms. 
Panels \textbf{D} and \textbf{E} show examples of narrow-band pulses, and \textbf{F} shows a pulse with notable frequency drift.
The single pulses are intermittent and far narrower than the integrated profiles, with a mean Gaussian width of 0.9~ms. This might suggest that the physical condition for pair production and radio emission is not satisfied globally in the magnetosphere but are rather temporarily satisfied in some local regions.
Most single pulses appear in a narrow band with the bandwidth peaking around 0.29~GHz. The narrow spectra differ from the flat spectra reported in some other magnetar transient radio emission\cite{camilo07}. 
The resemblance between the magnetar pulsar emission and repeating FRB pulses may be indicative of a connection between the two. However, similar features have also been observed from some canonical pulsars \cite{Bilous21}.

\subsubsection*{Pulse energy distribution}

The FAST has a gain of $\sim$16~ K~Jy$^{-1}$ \cite{Jiang20RAA} when fully illuminated, and the 19-beam L-band receiver has a system temperature of $T_{\rm sky}\sim23$~k, but these parameters vary slightly when the telescope is pointing at a high zenith angle ($>26.4^{\circ}$) and operating in a partially-illuminated mode \cite{Jiang20RAA}.
We computed the energies for every pulse based on the measured telescope performance parameters for the zenith angle of the telescope at the moment the pulse was observed.
The gain and system temperature could be derived for a given azimuth angle using the parameterized formula in Jiang et al. 2020\cite{Jiang20RAA}. 
The average systematic error in the derived gain across different frequencies is about 0.25~K~Jy$^{-1}$ (or 1.5\% relative to the average effective gain).

We conducted a simulation to estimate the completeness of our search procedure. 
We simulated pulses based on the best fit 2D Gaussian profiles from fitting the pulse shapes and injected these pulses into data that contained no detectable pulses.
The data injected with simulated pulses are pre-processed in the same fashion as the data used in our analysis, i.e., with polarisation channels merged, level adjusted, and 8-bit digitization. 
The 2D simulated pulses are also digitized in 8~bit before being injected into the 8~bit data.
When digitizing simulated pulses, we converted the power value in a pixel in the 2D profile from float to integer number and lose the fraction part. 
To make sure this does not cause a loss of signal strength, we generated a random number between 0 and 1, and when the random number was smaller than the fractional part of the float value, we gave the pixel an additional value of 1; otherwise, we kept only the integer part.
This way, we could simulate the effect of 8-bit digitization on detecting weak pulses.
We used seven one-hour observations that contained no detectable pulses and injected evenly one pulse per second with expected S/N in 5-12. We injected 20973 pulses and detected 7912 of them using our pipeline and eye inspections. 
Most injected signal falls into the lowest energy bin (9005 in total, of which only 1318 were detected), while the smallest number of injected signals in the other energy bins is 15.
Then we ran the simulated data through our detection pipeline and estimated the rate of correct detections for pulses with different S/N and energies. 

We used these simulations to correct our measured pulse energies distribution for potentially missed pulses. 
We analyzed the simulated pulses, computed their energy in the same way as for real detections, and binned them in the same grid as our measured energy distribution histogram.
We counted the number of detection versus the number of injections in our simulation to get the estimated detection rate and its propagated error for each bin.
We then divided the counts of the energy histogram by the detection rate to estimate the corrected histogram as in \EXTFIG{fig:energy}~\textbf{B}, where both the error in counts and detection rate are included in the error bars.
We fit the corrected pulse energy distribution with a log-normal function, and the best-fit central energy scale is $10^{26.45(1)}$~erg with a $\sigma=0.247(8)$ dex and a reduced $\chi^2$ of 0.7 for 7 degrees of freedom.

\subsubsection*{NICER X-ray observations}
We reduced the NICER data from epochs close to the radio burst detections using the HEAsoft-6.30.1 \cite{HEAsoft} and extracted soft X-ray events between 1-3~Kev.
We corrected the X-ray event times to the TDB scale at the barycenter of the solar system by using the \textsc{barycorr} command. 
We selected observations in days before and after the radio bursts and combined the barycentered events from these observations into one merged event list by using the \textsc{ftmerge} command.
For comparing the phases of FRB~20200428 and FAST \#1 (\FIG{fig:spds}~\textbf{B}) we used NICER observations between April 28 and 30.
We excluded the very first NICER good time interval from April 28 when the source was highly active and also excluded photons from X-ray bursts from the following intervals using the same method described in Ref. \cite{Younes20}. 
We fixed $\dot{P}$ to the our later measured value of 3.717$\times$10$^{-11}$ and found a best folding period of 3.247315(3)~s for the epoch 58968.90034 using H-test\cite{de2010h}, consistent with the best folding period found in the previous study \cite{borghese2020x, Younes20}. 
Because of the short time span, the maximum change in photon arrival phase is about 0.01 even if we allow $\dot{P}$ to change from 0 to double the assumed value.
When comparing the pulse profile of the persistent X-ray emission with the barycentered arrival time of FRB~20200428 and FAST \#1 (\FIG{fig:spds}~\textbf{B}), 
we found that FRB~20200428's radio phase seems to coincide with the peak of the X-ray profile, as seen by Ref. \cite{Younes20}, while the FAST \#1 arrives at the foot of the X-ray profile, similar to the pulses detected in our October 2020 campaign.   
For the Westerbork bursts, we used NICER data between May 13 and June 1.
Although using slightly different sets of data, we used the timing solution found by Ref. \cite{Younes20} for these epochs to fold the NICER events and resulting in a pulse profile that resembles those observed in other epochs. 
The relative phases found for the Westerbork bursts are consistent with those reported in Ref. \cite{Younes20}.
For the CHIME bursts, we used NICER data from October 6 to 28.
We used our radio timing solution (\TAB{tab:parfile}) to fold the NICER events and again got a pulse profile similar to those observed elsewhere. 
The CHIME burst phases do not seem to be coinciding with other bursts, and like the Westerbork bursts, the three CHIME bursts came in with broad phase separations.  
Finally, we also compared the integrated profile from the simultaneous observation by FAST and XMM-Newton taken on October 18, showing that the radio pulse came from the foot of the X-ray profile.

\subsubsection*{XMM-Newton observations}
We took two XMM-Newton observations, first one on 2020 Oct 18 for 9 hours from 03:53:16.775 to 13:14:18.424(UTC) simultaneously with FAST for 5 hours from 08:10:00 to 13:10:00(UTC), second one on 2020 Nov 13 for 11~hr from  02:23:33.379 to 13:26:12.775 (UTC).
The last 5 hours of the two observations are entirely overlapping.
We used SAS 18.0.0 and the \textsc{epchain} command to reduce the XMM-Newton data. 
We selected 5345 good events with standard criteria from the source using a circular region with a radius of 30" and energy between 1 and 3~keV.
We then corrected the timestamps of these events by using the \textit{barycorr} command in SAS,
and fold the corrected events into pulse profiles by using the timing ephemeris we get from Radio pulses with the \textit{prepfold} command in \textsc{PRESTO}.  
The folded profile has a reference MJD for its first bin, and this MJD is converted into a time of arrival and fed to \textsc{tempo}. 
We used \textsc{tempo} to predict the pulse phase of the X-ray reference MJD's phase while fixing the timing model to our best fit radio ephemeris. 
The resulting phase allows us to align the X-ray pulse profile and the Radio pulse profile, as shown in \FIG{fig:spds}~\textbf{B}. 
We folded the 2020-Oct-18 X-ray events into a 32-bin profile. 
Ref. \cite{Younes22} reported the root-mean-squared (RMS) pulsed fractions of 0.09(1) were estimated for the 2020 Oct 1 and 18 XMM-Newton 1-3~keV data based on the Fourier powers of the profiles.
Ref. \cite{Borghese22} reported (RMS) pulsed fraction of 0.13(1) for 1-2~keV XMM-Newton data on 2020 Oct 1.
Using the same definition, we also get a consistent $0.13(4)$ for the Oct 18 XMM-Newton data.

\subsubsection*{Hard X-ray observations}

During the radio activity of SGR J1935$+$2154 in October 2020, the {\it Insight}-HXMT\cite{HXMT} (Hard X-ray Modulation Telescope) performed five coordinated observations of the source with FAST on 10 October 8:00:00$-$11:18:14, 11 October 7:52:37$-$11:10:53, 13 October 7:38:02$-$12:28:12, 15 October 8:58:19$-$12:14:14 and 16 October 8:50:20$-$12:14:23 (all times are in UTC). All {\it Insight}-HXMT observations used three telescopes, namely, the low-energy X-ray telescope (LE; 1-10 keV), the medium-energy X-ray telescope (ME; 10-30 keV) and the high-energy X-ray telescope (HE; 27- 250 keV) covering 1-250 keV. No statistically significant burst was identified from the blind search based on the signal to noise ratio method \cite{cai2022}. Assuming the same duration (0.5 s) and spectral shape with the hard X-ray burst associated with FRB~20200428 detected by {\it Insight}-HXMT \cite{Li2021}, we calculated the $3\sigma$ upper limit on the burst fluence as $4.5 \times 10^{-9}$~ erg~cm$^{-2}$, $5.0 \times 10^{-9}$~erg~cm$^{-2}$ and $2.8 \times 10^{-9}$~erg~cm$^{-2}$ for LE, ME and HE, respectively.

\subsubsection*{Emission geometry constraints}

%In this 2020 October episode, SGR~J1935$+$2154 behaved like a pulsar, and its single pulses fell in a narrow phase range of 0.07.  
The FAST detection of single pulses indicates that the condition for coherent emission persisted on the monthly timescale in a localized region of SGR~J1935+2154's magnetosphere, albeit that pair production cascade, which is believed to be essential for radio emission, was triggered only intermittently. For a dipolar magnetic field, 
%the size and altitude of the emission region could be in general estimated from a dipolar magnetic field geometry,
the opening angle of the open field line region as a function of radius $r$ may be estimated as\cite{Rankin90} $\rho = 2.49^{\rm o} (r/R)^{1/2} P^{-1/2} / \sin\alpha$, where $P$ is the period, $\alpha$ is the inclination angle of the magnetic pole, and $R$ is the neutron star radius. In general, one should have \REV{emission open angle} $W \leq \rho$, since the line of sight may not cut through the center of the emission beam. Taking $P=3.2478$~s, one obtains a constraint $r/R \geq 240 \sin\alpha$. 
Noticing that the light cylinder radius is $r_{\rm lc}/R \sim 1.55 \times 10^4$ for the source. 
The inferred $r/R$ is a relatively high emission altitude, provided that the magnetic inclination angle is not too small ($\sin\alpha \ll 1$). 

\begin{figure}[H]
  \centering
  \includegraphics[width=5.0in]{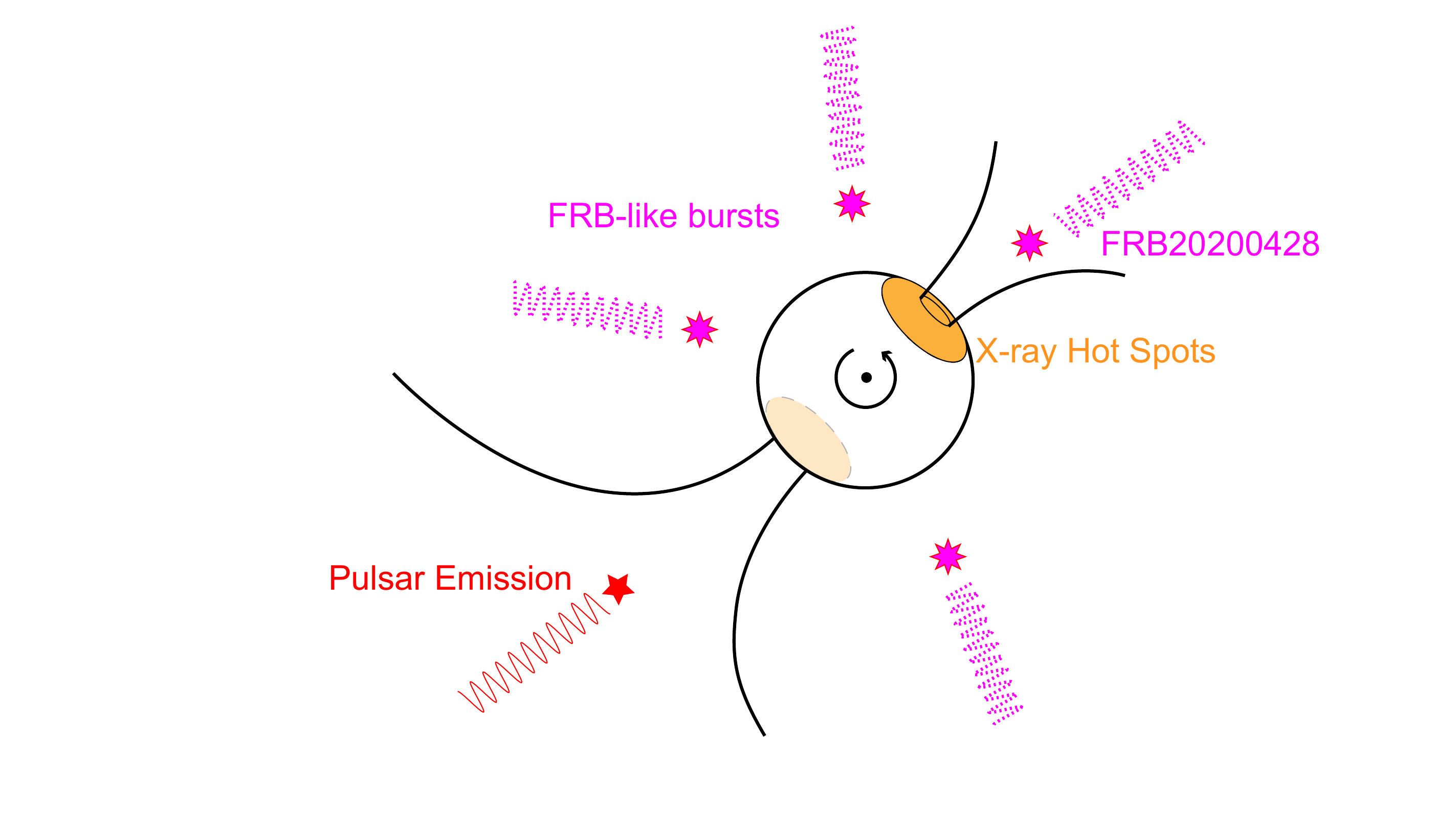}
  \caption{
\textbf{The illustration of a possible picture for SGR~J1935$+$2154's radio and X-ray emission geometry.}
The X-ray hot spot (orange region) lies on the opposite side of the pulsar emission zone (red star), leading to a $\simeq$180$^{\circ}$ phase difference between the X-ray peak and the Radio pulses.
We also illustrate in purple an emission geometry for FRB~20200428 deduced from the illustrated pulsar emission geometry assuming a dipolar magnetic field.  
Note this illustration is not to scale since that dipolar magnetic field line only starts to bend over at altitude $\gtrsim10^4$~$R_{\rm ns}$ for such a slowly rotating magnetar.
}
  \label{fig:geometry}
 \end{figure}

Additionally, the fact the radio pulsation is anti-aligned with the X-ray pulsation indicates that these pulses came from the opposite side from the FRB~20200428 event. Based on these, we could draw a crude picture for the emission geometry of SGR~J1935+2154's bursts and pulses (\EXTFIG{fig:geometry}).
\REV{Note that this picture illustrates one possible geometry that could explain the observations, and \EXTFIG{fig:geometry} is not meant to be taken to scale. }

\begin{deluxetable}{lc}
\tabletypesize{\footnotesize}
\tablewidth{0pt}
\tablecaption{ Timing model parameters.\label{tab:parfile} }

\tablehead{ \colhead{Parameter}  &\colhead{Value}   }
\startdata
\textit{Measured Parameters}&  \\
%Spin Frequency $\nu$~(s$^{-1}$)&  0.3078991894(7)\\
%Spin down rate $\nu'$ (s$^{-2}$)&  $-3.5235(8)\times10^{-12}$\\
Spin Frequency $\nu$~(s$^{-1}$)&  0.307899190(3)\\
Spin down rate $\dot{\nu}$ (s$^{-2}$)&  $-3.524(3)\times10^{-12}$\\
\\
\textit{Fixed Parameters}&  \\
Dispersion Measure (pc~cm$^{-3}$)&  332.703
\tablenotemark{a}\\
Reference epoch (MJD)&  59131.377113\\
Right Ascension, $\alpha$ (J2000)&  19:34:55.5978 \tablenotemark{b}\\
Declination, $\delta$ (J2000)&  21:53:47.7864 \tablenotemark{b}\\
Solar System Ephemeris&  DE436\\
\\
\textit{Derived Parameters}&  \\
Dipole magnetic field, $B$ (G)&  $3.52\times10^{14}$\\
Characteristic age, $\tau_c$ (yr)&  $1.38\times10^{3}$
\enddata
\tablenotetext{a}{fixed to the weighted average DM value of single pulses.}
\tablenotetext{b}{position taken from X-ray imaging\cite{israel16} with an uncertainty of 0.4".}
\end{deluxetable}

\bigskip
\bigskip
\bigskip
%\bibliography{FRB}
%Where the bibliography will be printed
  %\printbibliography

%\section*{Formatting Citations}
\subsubsection*{Acknowledgments}
This work made use of data from the FAST, a Chinese national mega-science facility, built and operated by the National Astronomical Observatories, Chinese Academy of Sciences. 
This work made use of the data from the XMM-Newton, an European Space Agency's mission, and {\it Insight}-HXMT, a project funded by the China National Space Administration (CNSA) and the Chinese Academy of Sciences (CAS). We gratefully appreciate the arrangement of the coordinated Target of Oppotunity observations with FAST by both XMM-Newton and {\it Insight}-HXMT teams. We appreciate the helpful discussion with Joseph D. Gelfand. We appreciate the National Astronomical Data Center of China for hosting the data used in this paper.

\subsubsection*{Funding}
The authors acknowledge support from the National SKA Program of China No. 2020SKA0120200, 2020SKA0120100 (WWZ, KJL, HX, DJZ, JRN, BJW, LQM, JCJ, CCM, JWX, MYX, MY)% SKA
the National Nature Science Foundation grant No. 12041303, 11873067, 12041304, 12003028, 12203045 (WWZ, LL, LQM, CCM, MYX, YF)
YSBR-063, CAS Project for Young Scientists in Basic Research (WWZ, PJ, CHN, XLM)
the National Key R$\&$D Program of China No. 2017YFA0402600, 2021YFA0718500 (YTC, PW, DL,YLY, YKZ, LL)
the CAS-MPG LEGACY project, and funding from the Max-Planck Partner Group (PW, DL, YLY, KJL)
China Manned Spaced Project CMS-CSST-2021-B11 (YPY).

\subsubsection*{Data and code availability}
Raw data are available from the FAST data center: \url{http://fast.bao.ac.cn}. Owing to the large data volume, please contact the corresponding author or FAST data center for the data transfer. 
Both XMM-Newton and Insight-HXMT data are public and can be downloaded from the archive : http://nxsa.esac.esa.int/nxsa-web/ and  http://archive.hxmt.cn/proposal, respectively. 
The directly related data that support the findings of this study can be found:\url{https://nadc.china-vo.org/res/r101221/} and 
\url{http://groups.bao.ac.cn/psr/xsdt/202206/t20220607_704493.html}
All data needed to evaluate the conclusions in the paper are present in the paper and/or the Supplementary Materials.

\noindent
\textsc{PSRCHIVE} (\url{http://psrchive.sourceforge.net})

\noindent
\textsc{PRESTO} (\url{https://github.com/scottransom/presto})

\noindent
\textsc{BEAR} (\url{https://psr.pku.edu.cn/index.php/publications/frb180301/})

\noindent
\textsc{TransientX}
(\url{https://github.com/ypmen/TransientX})

\noindent
\textsc{dspsr} (\url{http://dspsr.sourceforge.net})

\noindent
\textsc{tempo} (\url{http://tempo.sourceforge.net})

\noindent
\textsc{tempo2} (\url{https://bitbucket.org/psrsoft/tempo2})

\noindent
\textsc{psrfits\_utils}  (\url{https://github.com/demorest/psrfits\_utils})

\noindent
\textsc{SAS 18.0.0} (\url{https://www.cosmos.esa.int/web/xmm-newton/sas-download})

\noindent
\textsc{HEAsoft 6.30.1}
(\url{https://heasarc.gsfc.nasa.gov/docs/software/heasoft/})

\noindent
\textsc{hxmtsoft v2.05}
(\url{http://hxmten.ihep.ac.cn/software.jhtml})

\subsubsection*{Author Contributions}
 WWZ, KJL, BZ coordinated the radio observation campaign and led the paper writing. WWZ, HX, DJZ, KJL led the radio and X-ray data reduction and analysis. LL, EG, CK carried out the X-ray observation. BJW, CFZ, PW, JRN, YTC, CKL, LQM, CCM, XLM, MY, MYX, XG participated in the data reduction and analysis. YF, JCJ, CHN, JWX, YKZ, YPM, JLH, DL, PJ, RXX, MYG, WYW, ZW, YPY, YLY, SNZ, participated in discussions and the writing of the paper. 

\subsubsection*{Competing Interests}
 The authors declare that they have no competing interests.

%\subsubsection*{Correspondence}
%Correspondence and requests for materials
%should be addressed to W. W. Zhu (zhuww@nao.cas.cn), K. J. Lee (kjlee@pku.edu.cn), and B. Zhang (bing.zhang@unlv.edu).

\clearpage
\newpage

\setcounter{figure}{0}
\setcounter{table}{0}
\captionsetup[table]{name={\textbf{Table S}},  textfont={bf}}
\captionsetup[figure]{name={\textbf{Fig. S}},  textfont={bf}}

\section*{Supplementary Materials}

In \EXTTAB{tab:Obs}, we list the parameters of FAST observations from April 2020 to February 2021 and the number of pulses detected.

%\section*{Supplements}

%\clearpage
\begin{longtable}{cccc}
\caption{List of FAST observations of SGR J1935+2154 \label{tab:Obs}}\\
%\tablewidth{0pt}
%\tablehead{ \colhead{start \\ (MJD)}  &\colhead{UTC}  &\colhead{Duration \\ (s)}  &\colhead{pulse counts}   }
%\startdata
\hline
Start& UTC & Duration & pulse counts  \\
(MJD)&      & (s)      &    \\
\hline
58954.9128356&  2021-04-15 21:54:29&  7200&  0\\
58965.8798032&  2020-04-26 21:06:55&  7200&  0\\
58966.9965277&  2020-04-27 23:55:00&  3340&  0\\
58967.8576388&  2020-04-28 20:35:00&  10800&  0\\
58968.8947916&  2020-04-29 21:28:30&  3840&  0\\
58969.8900462&  2020-04-30 21:21:40&  3840&  1\\
58970.8750000&  2020-05-01 21:00:00&  3600&  0\\
58971.8778935&  2020-05-02 21:04:10&  3600&  0\\
58972.8933912&  2020-05-03 21:26:29&  3600&  0\\
58973.9375000&  2020-05-04 22:30:00&  5400&  0\\
58975.9375000&  2020-05-06 22:30:00&  3600&  0\\
58976.9375000&  2020-05-07 22:30:00&  5400&  0\\
58977.9375000&  2020-05-08 22:30:00&  5400&  0\\
58981.8946180&  2020-05-12 21:28:15&  4200&  0\\
58983.8750000&  2020-05-14 21:00:00&  5400&  0\\
58985.8761458&  2020-05-16 21:01:39&  5400&  0\\
58986.7947569&  2020-05-17 19:04:27&  3600&  0\\
58988.8737962&  2020-05-19 20:58:16&  8100&  0\\
59066.6774189&  2020-08-05 16:15:29&  3600&  0\\
59089.6402777&  2020-08-28 15:22:00&  3600&  0\\
59131.3812499&  2020-10-09 09:09:00&  3600&  86\\
59132.3645833&  2020-10-10 08:45:00&  3600&  150\\
59133.4066203&  2020-10-11 09:45:32&  3600&  204\\
59135.4211921&  2020-10-13 10:06:31&  3600&  5\\
59137.4583333&  2020-10-15 11:00:00&  3600&  16\\
59138.4465277&  2020-10-16 10:43:00&  3600&  11\\
59140.3402777&  2020-10-18 08:10:00&  18000&  85\\
59143.5347222&  2020-10-21 12:50:00&  1800&  7\\
59146.4056597&  2020-10-24 09:44:09&  3600&  30\\
59147.4888888&  2020-10-25 11:44:00&  3600&  24\\
59148.4944444&  2020-10-26 11:52:00&  3600&  26\\
59151.4430671&  2020-10-29 10:38:01&  3600&  92\\
59152.4517245&  2020-10-30 10:50:29&  3600&  59\\
59158.2986111&  2020-11-05 07:10:00&  3600&  0\\
59159.4576388&  2020-11-06 10:59:00&  3600&  0\\
59160.4597222&  2020-11-07 11:02:00&  3600&  0\\
59178.3930555&  2020-11-25 09:26:00&  3600&  0\\
59204.2083333&  2020-12-21 05:00:00&  3600&  0\\
59243.0694444&  2021-01-29 01:40:00&  3600&  0\\
59249.1527777&  2021-02-04 03:40:00&  3600&  0\\
%\enddata
\hline
\end{longtable}

\clearpage 

\end{document}